\documentclass[prd,aps,showpacs,nofootinbib,tightenlines]{revtex4}
\usepackage{amsmath}
\usepackage{amssymb}
\usepackage{epsfig}
\usepackage{graphicx}
\begin{document}

\title{Resonant state $D_0^\ast(2400)$ in the quasi-two-body $B$ meson decays}
\author{Wen-Fei Wang$^{1,2}$}\email{wfwang@sxu.edu.cn}
\affiliation{$^1$Institute of Theoretical Physics, Shanxi University, Taiyuan, Shanxi 030006, China} 
\affiliation{$^2$State Key Laboratory of Quantum Optics and Quantum Optics Devices, Shanxi University, 
                Taiyuan, Shanxi 030006, China}
\date{\today}

\begin{abstract}
We study four quasi-two-body decay processes including $D^\ast_0(2400)$ as the intermediate state in the perturbative QCD (PQCD) 
approach. The branching fraction predicted in this work for the decay mode $B^-\to D^{*}_0(2400)^0\pi^-\to D^+\pi^-\pi^-$ agree  
with the data from Belle, BaBar and LHCb Collaborations. The PQCD prediction of the branching ratio for the decay 
$\bar B^0\to D^{*+}_0K^-\to D^0\pi^+K^-$ is consistent with the value given by LHCb.
For the decays $\bar B^0\to D^{*+}_0\pi^-\to D^0\pi^+\pi^-$ and $B^-\to D^{*0}_0K^-\to D^+\pi^-K^-$, 
the PQCD predicted branch ratios are
$2.85^{+1.23}_{-0.80} (\omega_B) ^{+1.05}_{-0.81}(\omega_{D\pi})^{+0.33}_{-0.31}(a_{D\pi})
                  ^{+0.06}_{-0.05}(\Gamma_{D^{*+}_0}) \times 10^{-4}$ and 
$4.65^{+1.89}_{-1.30} (\omega_B) ^{+1.51}_{-1.24}(\omega_{D\pi})^{+0.40}_{-0.38}(a_{D\pi})
                  ^{+0.22}_{-0.18}(\Gamma_{D^{*0}_0}) \times 10^{-5}$, respectively.
We analyze the experimental branching fractions using the ratios $R_{D^{*0}_0}$ and $R_{D^{*+}_0}$ which are related to the 
decays with the neutral and charged $D^\ast_0(2400)$, respectively.
The available experimental results for the quasi-two-body decays including $D^\ast_0(2400)$ are not in agreement with the 
isospin relation and $SU(3)$ flavor symmetry. 
\end{abstract}

\pacs{13.20.He, 13.25.Hw, 13.30.Eg}
\maketitle
Many excited open-charm states have been discovered by various experiments in recent years, 
see Ref.~\cite{rpp80-076201} for a review.
One of them, the $p$-wave orbitally excited state $D^*_0(2400)$, with the light degree of freedom 
$j^p_q=\frac12^+$~\cite{plb572-164,prd72-054029,prd93-034035} and quantum number $J^P=0^+$~\cite{PDG-2018}, 
was first discovered by Belle Collaboration in the three-body decays $B^-\to D^+\pi^-\pi^-$, 
with the mass $m_{D^{*0}_0}=2308\pm17\pm15\pm28$ MeV and 
width $\Gamma_{D^{*0}_0}=276\pm21\pm18\pm60$ MeV~\cite{prd69-112002}.
For simplicity, we adopt $D^*_0$ to denote the $D^*_0(2400)$ state and the inclusion of charge-conjugate processes is 
implied throughout this work. The neutral resonant state $D^{*0}_0$ has been confirmed by BaBar Collaboration in the same 
decay processes in~\cite{prd79-112004}, with the close but preciser values for its mass and width.  
While in the wideband photoproduction experiment, differ from those three-body  decay processes, FOCUS Collaboration 
provided quite different values for the broad structure ${D^{*0}_0}$ in~\cite{plb586-11}.  One has 
$m_{D^{*0}_0}=2407\pm21\pm35$ MeV and $\Gamma_{D^{*0}_0}=240\pm55\pm59$ MeV  in company with 
$m_{D^{*+}_0}=2403\pm14\pm35$ MeV and $\Gamma_{D^{*+}_0}=283\pm24\pm34$ MeV for a charged state 
$D^{*+}_0$ from Ref.~\cite{plb586-11}. 

Unlike the charmed-strange state $D^*_{s0}(2317)$~\cite{prl90-242001,prd68-032002,prl91-262002}, 
which lies just below $DK$ threshold and mainly decays into the isospin breaking channel $D_s\pi$,  
the state $D^*_0$ is expected to decay rapidly through the $s$-wave pion emission, the conservation of its angular momentum
implies this resonance primarily couple to $D\pi$ and has a broad decay width~\cite{pr429-243} as revealed by experiments. 
While the discrepancy of its properties between the experimental results~\cite{prd69-112002,prd79-112004} and the predictions 
from the quark model~\cite{prd32-189,prd43-1679} has triggered many studies on its true nature.
The strong decays, radiative decays and/or the spectra have been studied extensively in Refs.~\cite{prd72-054029,prd72-094004,prd78-014029,prd87-034501,prd92-074011,prd93-034035,epjc66-197,epjc77-312,epjc78-583} 
to explore the exact feature of the resonant state $D^*_0$. 
In Refs.~\cite{plb624-217,npb161-193}, the possible four-quark structure of $D^*_0$ was  investigated, 
the authors pointed out that the four-quark structure is acceptable for the resonant state observed by 
 Belle~\cite{prd69-112002} and BaBar~\cite{prd79-112004}, but not for the cases observed by FOCUS~\cite{plb586-11}. 
While in Refs.~\cite{plb641-278,plb767-465,rmp90-015004}, it was claimed that there exist two poles in $D^*_0$ energy 
region. And a pole near the $D\pi$ threshold was obtained from lattice QCD in~\cite{jhep1610-011}, which was said to 
share the similarities with the experimental resonance $D^*_0$.  The resonant state $D^*_0$ has also been explained as a 
mixture of two- and four-quark state~\cite{prd73-034002} or the  bound state of $D\pi$~\cite{prd76-074016}.

Semileptonic or hadronic $B$ meson decays including a resonant state $D^*_0$ shall yield clues to its properties.
Employing constituent quark model or the light-cone sum rules to evaluate the $B\to D^*_0$ transition form factors, 
the decays of $B\to D^*_0l\nu$ have been studied in Refs.~\cite{epjc48-441,prd90-114015}.
With the help of a chiral unitarity model, the ratio between the decay widths of $\bar B^0_s\to D^{*+}_{s0}(2317)l\bar\nu_l$ and
$\bar B^0\to D^{*+}_{0}l\bar\nu_l$ was calculated in~\cite{prd92-014031}.
The model independent studies of $B\to D^{**}l\nu$ have been performed within the standard model in~\cite{prd95-014022}  and
beyond the standard model in~\cite{prd97-075011} based on heavy quark symmetry.
In~\cite{1808-02432}, the branching fraction for the semileptonic decay $B\to D^\ast_0 l\nu$ was predicted 
assuming the conventional quark-antiquark configuration for $D^\ast_0$ state.
And the hadronic matrix elements were evaluated in the Bethe-Salpeter approach for  $B_c\to D^*_0\mu\bar\mu$ decays 
in~\cite{jhep1509-171}. Since Belle's announcement~\cite{prd69-112002}, many works have been done on two-body hadronic 
$B$ meson decays including the state $D^*_0$. For example, within the covariant light-front approach, the branching ratio of 
$B^-\to D^{\ast0}_0\pi^-$ was predicted to be $7.3\times 10^{-4}$ in~\cite{prd69-074025}.
In~\cite{prd72-094010}, the information on the Isgur-Wise functions at zero recoil was extracted from the 
$\bar B^0\to D^{\ast+}_0\pi^-$ decay process. Using the improved version of the Isgur-Scora-Grinstein-Wise quark model for 
the $B\to D^*_0$ transition form factors, the decays $B^-\to D^{\ast0}_0\pi^-$ and $B^0\to D^{\ast-}_0\pi^+$ 
have their branching ratios as $7.7\times10^{-4}$ and $2.6\times10^{-4}$, respectively, in Ref.~\cite{prd68-094005}. 
Two-body decays $B\to  D^*_0\pi$ have also been discussed within factorization framework~\cite{prd74-034020}
and the perturbative QCD (PQCD) approach~\cite{prd68-114008}.

\begin{figure}[tbp]
\centerline{\epsfxsize=12cm \epsffile{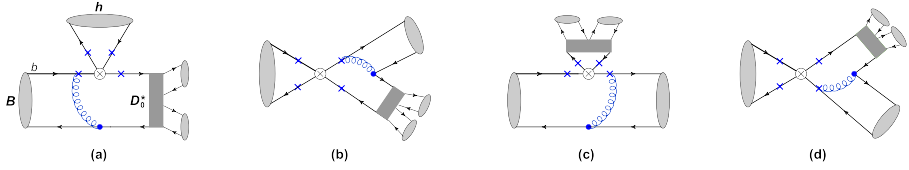}}
\caption{Typical Feynman diagrams for the quasi-two-body decays $B\to D^*_0 h\to D\pi h$, $h=$($\pi, K$). The symbol $\otimes$ 
stands for the weak vertex, $\times$ denotes possible attachments of hard gluons, and the grey rectangle represents the intermediate 
states $D^*_0$}
\label{fig-feyndiag}
\vspace{-0.3cm}
\end{figure}

\begin{table}[thb]
\begin{center}
\caption{Data for the quasi-two-body hadronic $B$ meson decays involving the $D^*_0$ as the resonant state}
\label{tab1}   
\begin{tabular}{l c l l} \hline
   ~~~Mode       &    ~Unit~      &   \;\;\;\;Branching fraction     &      ~~ Ref. ~ \\  \hline
  $B^-\to D^{*0}_0\pi^-\;\to D^+\pi^-\pi^-$\;     &$~~~(10^{-4})~~~$
      &\; $6.1\pm0.6\pm0.9\pm1.6 $\;  & Belle\;\;\;\cite{prd69-112002} \\
     ~    &$~~~(10^{-4})~~~$ &\; $6.8\pm0.3\pm0.4\pm2.0 $\;  & BaBar\;\cite{prd79-112004} \\
     ~    &$~~~(10^{-4})~~~$ &\; $5.78\pm0.08\pm0.06\pm0.09\pm0.39\footnotemark[1] $\;\;\;\;  & LHCb\;\cite{prd94-072001} \\
   $\bar B^0\;\to D^{*+}_0\pi^-\;\to D^0\pi^+\pi^-$\;     &$~~~(10^{-4})~~~$
      &\; $0.60\pm0.13\pm0.15\pm0.22 $\; & Belle\;\;\;\cite{prd76-012006} \\ 
     ~    &$~~~(10^{-5})~~~$ &\; $7.7\pm0.5\pm0.3\pm0.3\pm0.4\footnotemark[2] $\;\;\;\;  & LHCb\;\cite{prd92-032002} \\
     ~    &$~~~(10^{-5})~~~$ &\; $8.0\pm0.5\pm0.8\pm0.4\pm0.4\footnotemark[3] $\;\;\;\;  & LHCb\;\cite{prd92-032002}\\
  $B^-\to D^{*0}_0K^-\;\to D^+\pi^-K^-$\;     &$~~~(10^{-6})~~~$
      &\; $6.1\pm1.9\pm0.5\pm1.4\pm0.4 $\;  & LHCb\;\cite{prd91-092002} \\
  $ B^0\;\to D^{*-}_0K^+\to\bar D^0\pi^-K^+$\;     &$~~~(10^{-5})~~~$
      &\; $1.77\pm0.26\pm0.19\pm0.67\pm0.20 $\; & LHCb\;\cite{prd92-012012} \\ 
\hline
\end{tabular}
\footnotetext[1]{ Total  $S$-wave $D^+\pi^-$ contribution }
\footnotetext[2]{ Isobar model }
\footnotetext[3]{ K-matrix model }
\end{center}
\end{table}

In Table~\ref{tab1}, we have the data for the four quasi-two-body hadronic $B$ meson decay modes including $D^*_0$
from Belle, BaBar and LHCb Collaborations. 
In the processes $B\to D^*_0 h\to D\pi h$, where $h$ is a charged pion or kaon, the weak interaction point
accompany the birth of the bachelor particle $h$, the intermediate state $D^{*0(+)}_0$ generated from the hadronization 
of $c$-quark plus $\bar u(\bar d)$-quark as demonstrated in Fig.~\ref{fig-feyndiag}. 
We stress that the resonance $D^*_0$ is not necessary to be conventional quart-antiquark structure.
In this letter, we shall analyze those four decay processes in a quasi-two-body framework based on the PQCD factorization
approach~\cite{plb504-6,prd63-054008,prd63-074009,ppnp51-85}. In Refs.~\cite{plb561-258,prd70-054006,
prd89-074031,prd91-094024,prd97-034033}, the PQCD approach has been employed in the studies 
of the three-body $B$ meson decays. With the help of the two-pion distribution amplitudes~\cite{Mueller:1998fv,
Diehl:1998dk,npb555-231} and the experimental inputs for the time-like pion form factors, in Ref.~\cite{plb763-29}, we 
calculated the decays $B\to K\rho(770),K\rho^\prime(1450)\to K\pi\pi$ in the quasi-two-body framework. The method used 
in~\cite{plb763-29} have been adopted for some other quasi-two-body $B$ decays in Refs.~\cite{paps-li-ya,paps-li-ya-II,paps-ma-aj}.
In this work, we extend the previous studies to the $\bar B^{0}\to D^{*+}_0 h^-\to  D^0\pi^+h^-$ and
$B^{-}\to  D^{*0}_0 h^-\to D^+\pi^-h^-$ decays.

Refer to the $K\pi$ system in Refs.~\cite{npb622-279,prd80-054007}, we define the scalar form factor $F^{D\pi}_0(s)$ 
for the final state $D^+\pi^-$ decays from ${D}^{*0}_0$ as
\begin{eqnarray}
\langle D^+\pi^- | \bar c u| 0 \rangle = \sqrt2 B_0 F_0^{D\pi}(s)\;,
\end{eqnarray}
with the constant
\begin{eqnarray}
B_0 = \frac{m^2_D-m^2_\pi}{\sqrt2 (m_c-m_u)}\approx 1.93~{\rm GeV}\;,
\end{eqnarray}
where the $m_D$($m_\pi$) is the mass of $D$($\pi$) meson,
the $m_c=1.275$ GeV and $m_u=2.2$ MeV (which could be neglected safely) for the mass of $c$ and $u$ quarks are adopted 
from~\cite{PDG-2018}. Then we have
\begin{eqnarray}
\langle D^+\pi^- | \bar c u| 0 \rangle \approx  \langle D^+\pi^- |  D_0^{*0} \rangle \frac{1}{\mathcal{D}_{\rm BW}} 
 \langle  D_0^{*0} |\bar c u |0 \rangle = \Pi^{\rm BW}_{D_0^*D\pi}  \langle  D_0^{*0} |\bar c u |0 \rangle       \;,
\end{eqnarray}
and
\begin{eqnarray}
\Pi^{\rm BW}_{D_0^*D\pi}=\frac{g_{D_0^*D\pi}}{\mathcal{D}_{\rm BW}} 
=\frac{\sqrt2 B_0 F_0^{D\pi}(s)}{\langle  D_0^{*0} | \bar c u | 0 \rangle }
= \frac{\sqrt2 B_0}{\bar f_{D_0^*} m_0 } F_0^{D\pi}(s)\;,
\end{eqnarray}
with $\bar f_{D_0^*}= \frac{m_0}{m_c(\mu)-m_u(\mu)}\cdot f_{D_0^*}$, and $f_{D_0^*}$ is the decay constant of $D_0^{*}$. 
One has different values from $78$ MeV~\cite{prd86-094031} to $148^{+40}_{-46}$ 
MeV~\cite{prd74-034020} in different works for this decay constant, 
see~\cite{prd54-6803,plb650-15,jpg39-025005,prd86-094031,prd69-074025,prd74-034020,prd68-114008}, 
we support the moderate one $f_{D_0^*}=0.13$ GeV which was adopted in the PQCD approach in~\cite{prd68-114008}. 
The denominator ${\mathcal{D}_{\rm BW}} = m^2_0-s-i m_0\Gamma(s)$, the mass-dependent decay width $\Gamma(s)$ has
its definition as $\Gamma(s)=\Gamma_0\frac{q}{q_0}\frac{m_0}{\sqrt s}$, $m_0$ and $\Gamma_0$ are the pole mass and width 
of the resonant state $D_0^{*}$ and $s$ is the invariant mass square for the $D\pi$ pair in the final state. In the rest frame of the 
resonance $D_0^{*}$, its daughter $D^+$ or $\pi^-$ has the magnitude of the momentum as
\begin{eqnarray}
q=\frac{1}{2}\sqrt{\left[s-(m_D+m_\pi)^2\right]\left[s-(m_D-m_\pi)^2\right]/s}\;,
\end{eqnarray}
and $q_0$ is the value of $q$ at $s=m^2_0$. The coupling constant $g_{D_0^*D\pi}$ has its value 
from the relation~\cite{prd92-014031,prd68-094005}
\begin{eqnarray}
g_{D_0^*D\pi}=\sqrt{\frac{8\pi m^2_0 \Gamma_0}{q_0}}\;.
\end{eqnarray}
We define 
\begin{eqnarray}
F_{D\pi}(s) = \frac{m^2_0}{m^2_0-s-i m_0\Gamma(s)}\;,
\end{eqnarray}
then we have $F_0^{D\pi}(s)=C_{D\pi}\cdot F_{D\pi}(s) $, with the parameter 
\begin{eqnarray}
C_{D\pi} = \frac{g_{D_0^*D\pi}\bar f_{D_0^*} }{\sqrt2 B_0 m_0} 
\end{eqnarray}

In the rest frame of the $B$ meson, with $m_B$ being its mass, we define the momentum  $p=\frac{m_B}{\sqrt 2}(1, \eta,0)$  
in the light-cone coordinates for the resonant state $D^*_0$ and the $D\pi$ pair coming out from the resonance. Its easy to see 
$\eta=s/m_B^2$ with $s=p^2$. The light spectator quark comes from $B$ meson and goes into $D^*_0$ in the hadronization 
processes in Fig.~\ref{fig-feyndiag}~(a) 
got the momentum $k=(\frac{m_B}{\sqrt 2}z, 0, k_{\rm T})$, $z$ is the momentum fraction. The momenta 
$p_B, p_3, k_B$ and $k_3$ for the $B$ meson, bachelor meson $h$ and the associated spectator quarks for $B$ and $h$ have their
definitions as
\begin{eqnarray}
p_B=\frac{m_B}{\sqrt2}(1,1,0_{\rm T}),\quad
p_3=\frac{m_B}{\sqrt2}(0,1-\eta,0_{\rm T}),\quad
k_B=\left(0,\frac{m_B}{\sqrt2}x_B ,k_{B{\rm T}}\right),\quad
k_3=\left(0,\frac{m_B}{\sqrt2}(1-\eta)x_3,k_{3{\rm T}}\right),
\end{eqnarray}
where $x_B$ and $x_3$ are the corresponding momentum fractions.

The $S$-wave $D\pi$ system distribution amplitude could be collected into~\cite{prd91-094024,prd68-114008,plb763-29}
\begin{eqnarray}
\Phi^{S-wave}_{D\pi}=\frac{1}{\sqrt{2N_c}}\left({ p \hspace{-1.6truemm}/ } +\sqrt s\right)C_{D\pi} \phi_{D\pi}(z,b,s)\;,
\end{eqnarray}
and the distribution amplitude
\begin{eqnarray}
\phi_{D\pi}(z,b,s)=\frac{F_{D\pi}(s) }{2\sqrt{2N_c}}\left\{ 6z(1-z)\left[\frac{m_c(s)-m_u(s)}{\sqrt s}+a_{D\pi}(1-2z) \right]\right\}
{\rm exp}\left(-\omega^2_{D\pi}b^2/2 \right)\;,
\label{def-wavefun}
\end{eqnarray}
the $a_{D\pi}=0.40\pm0.10$ and $\omega_{D\pi}=0.40\pm0.10$ GeV are adopted in the calculation in this work
by catering to our numerical results to the data in Table~\ref{tab1} and considering the related parameters for
the $D$ and $D^*$ mesons in the literature. The numbers for $\omega_{D\pi}$ and $a_{D\pi}$ in Ref.~\cite{prd68-114008} have 
been considered as the references in this work, but we don't use the same values because of the different 
framework of the two-body and quasi-two-body decays and the different definitions of the distribution amplitudes. 
The distribution amplitudes for the pion, kaon and $B$ meson are the same as those widely adopted in the 
PQCD approach to hadronic B meson decays, one can find their expressions and the relevant parameters in 
Ref.~\cite{prd86-114025}.

The decay amplitude ${\mathcal A}$ for the quasi-two-body decay processes 
$\bar B^{0}\to D^{*+}_0 h^-\to  D^0\pi^+h^-$  and $B^{-}\to  D^{*0}_0 h^-\to D^+\pi^-h^-$ in the PQCD approach 
is given by~\cite{plb561-258,prd70-054006,plb763-29}
\begin{eqnarray}
{\mathcal A}=\phi_B\otimes H\otimes\phi_{h}\otimes \phi_{D\pi}\;,
\end{eqnarray}
where the symbol $\otimes$ means convolutions in parton momenta, the hard kernel $H$ contains only one hard gluon exchange 
at leading order in the strong coupling $\alpha_s$ as in the two-body formalism and the distribution amplitude 
$\phi_B$ ($\phi_{h}$, $\phi_{D\pi}$) absorbs nonperturbative dynamics in the decay processes. 
We then have the differential branching fraction (${\mathcal B}$)~\cite{PDG-2018}
\begin{eqnarray}
\frac{d\mathcal{B}}{d\eta}=\tau_B \frac{q_h q B^2_0 C^2_{D\pi} }
 {32 \pi^3  m_B m^2_0 }\overline{|{\cal A}|^2} \;,\label{eqn-diff-bra}
\end{eqnarray}
$\tau_B$ being the $B$ meson mean lifetime, the magnitude momentum for bachelor $h$, in the center-of-mass 
frame of the $D\pi$ pair, as
\begin{eqnarray}
q_h=\frac{1}{2}\sqrt{\big[\left(m^2_{B}-m_{h}^2\right)^2 -2\left(m^2_{B}+m_{h}^2\right)s+s^2\big]/s}\;.
\end{eqnarray}
The $m_h$ is the mass of the bachelor meson pion or kaon.

In the numerical calculation, we adopt $\Lambda^{(f=4)}_{ \overline{MS} }=0.25$~GeV.  The decay constant $f_B=0.19$ GeV for 
$B$ meson  comes from lattice QCD~\cite{1712-09262}. The masses and the mean lifetimes for the neutral and 
charged $B$ meson,  the pole masses and the widths of the neutral and charged $D^*_0$ state, the Wolfenstein 
parameters, the masses of pion, kaon and D meson are all come from the Particle Data Group~\cite{PDG-2018}.
Utilizing the the differential branching fraction Eq.~(\ref{eqn-diff-bra}) and the decay amplitudes collected in Appendix A, we obtain
the branching fractions in Table~\ref{tab2} for the concerned quasi-two-body decay processes.
The shape parameter uncertainty of the $B$ meson, $\omega_{B}=0.40\pm0.04$ GeV, contributes the largest error for the branching 
fractions in Table~\ref{tab2}, the $\omega_{D\pi}=0.40\pm0.10$ GeV for $D\pi$ system takes the second place 
and the $a_{D\pi}=0.40\pm0.10$ in the Eq.~(\ref{def-wavefun}) generates the third one. For the decay width of the resonance $D^{*}_0$, 
the charged state got $\Gamma_{D^{*+}_0}=230\pm17$ MeV and neutral one has $\Gamma_{D^{*0}_0}=267\pm40$ 
MeV~\cite{PDG-2018}, then we have the quite different weight of the error from decay width for those processes including charged 
and neutral $D^{*}_0$ state, as shown in Table~\ref{tab2}. There are other errors, which come from the uncertainties of the 
parameters in the distribution amplitudes for bachelor pion(kaon)~\cite{prd86-114025} and the Wolfenstein 
parameters~\cite{PDG-2018}, are small and have been neglected.

\begin{table}[thb]
\begin{center}
\caption{PQCD predictions for the concerned quasi-two-body decays involving the $D^*_0$ as the intermediate state}
\label{tab2}   
\begin{tabular}{l c l } \hline
    ~~~Mode       &    ~Unit~      &   \;\;\;\;Branching fraction \\  \hline
  $B^-\to D^{*0}_0\pi^-\;\;\to D^+\pi^-\pi^-$\;     &$~~~(10^{-4})~~~$
      &\; $5.95^{+2.37}_{-1.64}(\omega_B) ^{+1.97}_{-1.55}(\omega_{D\pi})^{+0.54}_{-0.49}(a_{D\pi})
                  ^{+0.29}_{-0.21}(\Gamma_{D^{*0}_0})$\;  \\
   $\bar B^0\;\to D^{*+}_0\pi^-\;\to D^0\pi^+\pi^-$\;     &$~~~(10^{-4})~~~$
      &\; $2.85^{+1.23}_{-0.80}(\omega_B) ^{+1.05}_{-0.81}(\omega_{D\pi})^{+0.33}_{-0.31}(a_{D\pi})
                  ^{+0.06}_{-0.05}(\Gamma_{D^{*+}_0}) $\;  \\ 
  $B^-\to D^{*0}_0K^-\;\to D^+\pi^-K^-$\;     &$~~~(10^{-5})~~~$
      &\; $4.65^{+1.89}_{-1.30}(\omega_B) ^{+1.51}_{-1.24}(\omega_{D\pi})^{+0.40}_{-0.38}(a_{D\pi})
                  ^{+0.22}_{-0.18}(\Gamma_{D^{*0}_0})  $\;  \\
  $\bar B^0\;\to D^{*+}_0K^-\to D^0\pi^+K^-$\;     &$~~~(10^{-5})~~~$
      &\; $2.38^{+0.95}_{-0.65}(\omega_B) ^{+0.85}_{-0.68}(\omega_{D\pi})^{+0.30}_{-0.28}(a_{D\pi})
                  ^{+0.04}_{-0.03}(\Gamma_{D^{*+}_0})  $\;  \\ 
\hline
\end{tabular}
\end{center}
\vspace{-0.2cm}
\end{table}

\begin{figure}[tbp]
\centerline{\epsfxsize=7.0cm \epsffile{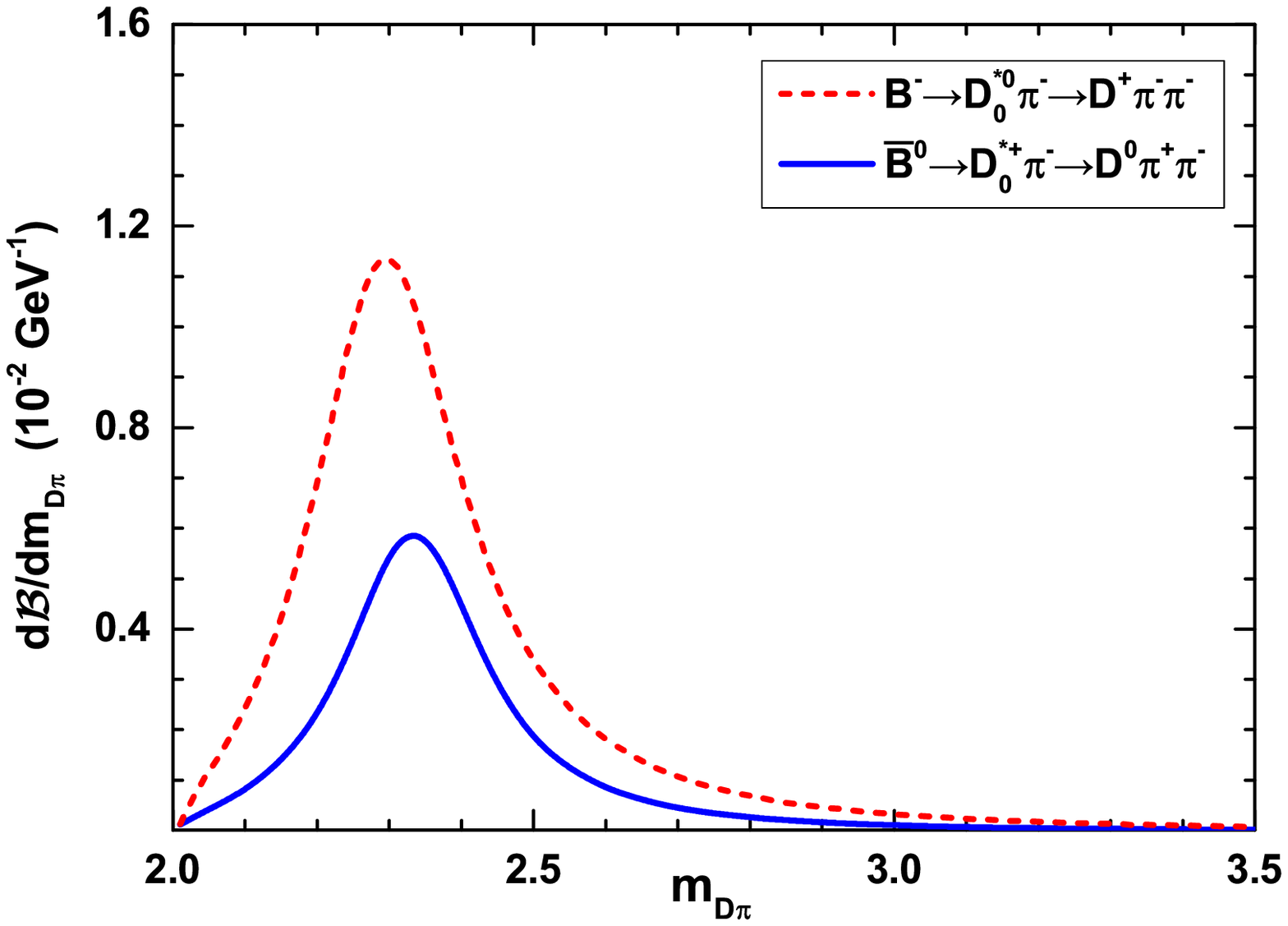}
                  \epsfxsize=7.0cm \epsffile{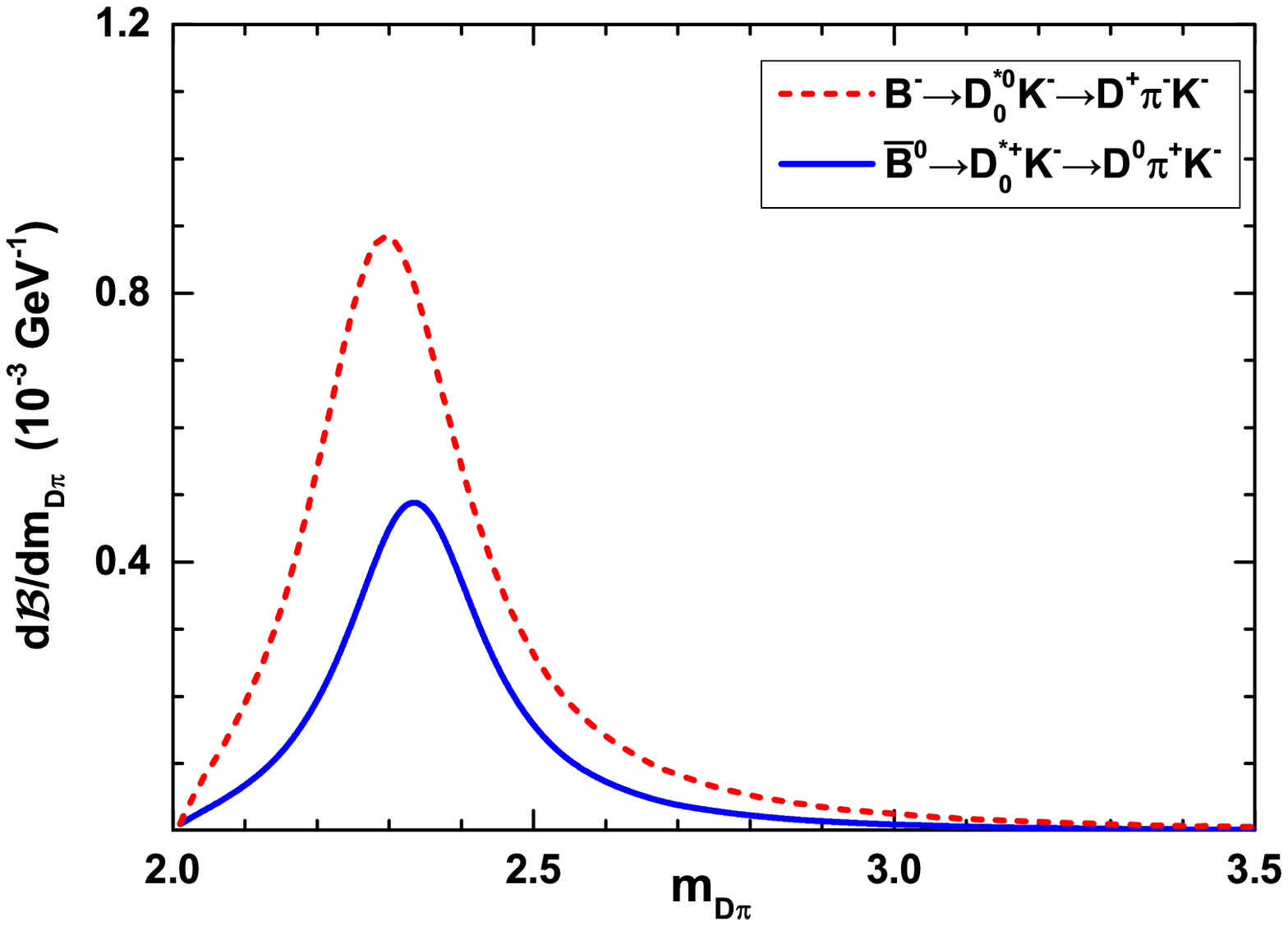}}
\vspace{-0.5cm}
\caption{The differential branching fractions for the decays $B^-\to D^{*0}_0\pi^-\to D^+\pi^-\pi^-$ and 
$\bar B^0\to D^{*+}_0\pi^-\to D^0\pi^+\pi^-$ (left), $B^-\to D^{*0}_0K^-\to D^+\pi^-K^-$ and  
$\bar B^0\to D^{*+}_0K^-\to D^0\pi^+K^-$ (right)}
\label{fig-deps}
\vspace{-0.2cm}
\end{figure}

The distributions of those four branching ratios in Table~\ref{tab2} in the $D\pi$ pair invariant mass $m_{D\pi}$
are shown in Fig.~\ref{fig-deps}, with the curves for $B^-\to D^{*0}_0\pi^-\to D^+\pi^-\pi^-$ (the dash line) and 
$\bar B^0\to D^{*+}_0\pi^-\to D^0\pi^+\pi^-$ (the solid line) on the left,  and the curves 
for  $B^-\to D^{*0}_0K^-\to D^+\pi^-K^-$ (the dash line) and  $\bar B^0\to D^{*+}_0K^-\to D^0\pi^+K^-$
(the solid line) at the right. The small mass difference of the charged and the neutral $D^*_0$ exhibit the different peaks of 
the $m_{D\pi}$ dependence for the different decay modes. The main portion of each branching ratio lies obviously in the 
region around the pole mass of the resonant state $D^*_0$ in the Fig.~\ref{fig-deps}, the contributions from the energy region 
$m_{D\pi}>3$ GeV can be safely omitted.

Assuming the $D^*_0$ state decays essentially into two-body modes, from the Clebsch-Gordan coefficients, we have
${\mathcal B}(D^{*0}_0\to D^+\pi^-)={\mathcal B}(D^{*+}_0\to D^0\pi^+)=\frac23$. Then we have the two-body results as
${\mathcal B}(B^-\to D^{*0}_0\pi^-)=8.93^{+4.71}_{-3.48}\times 10^{-4}$ and 
${\mathcal B}(\bar B^0\to D^{*+}_0\pi^-)=4.28^{+2.48}_{-1.77}\times 10^{-4}$ from Table~\ref{tab2}.
The two-body value for the $B^-\to D^{*0}_0\pi^-$ decay agree well with the results in 
Refs.~\cite{prd69-074025,prd72-094010,prd68-094005,prd68-114008}. The result $4.28^{+2.48}_{-1.77}\times 10^{-4}$ for the decay
$B^-\to D^{*0}_0\pi^-$ is consistent with the results $2.6\times10^{-4}$ in~\cite{prd68-094005} and $3.1\times10^{-4}$ 
in~\cite{prd74-034020} within errors, but it's smaller than the corresponding results in Ref.~\cite{prd68-114008}.

Compare our numerical results in Table~\ref{tab2} with the corresponding data in Table~\ref{tab1}, we find that the PQCD 
value of the branching fraction 
for the quasi-two-body decay process $B^-\to D^{*0}_0\pi^-\to D^+\pi^-\pi^-$ in this work agree well with the values
$(6.1\pm0.6\pm0.9\pm1.6)\times10^{-4}$ taken from Belle~\cite{prd69-112002} and $(6.8\pm0.3\pm0.4\pm2.0)\times10^{-4}$ 
picked up from BaBar~\cite{prd79-112004}. In Ref.~\cite{prd94-072001}, LHCb Collaboration presented a result 
 $(5.78\pm0.08\pm0.06\pm0.09\pm0.39)\times10^{-4}$ for the total $S$-wave $D\pi$ system, 
which should be supposed to mainly contributed by the $D^*_0$ state, in the $B^-\to D^+\pi^-\pi^-$ decays. 
For the decay $\bar B^0\to D^{*+}_0K^-\to D^0\pi^+K^-$, the result 
$2.38^{+0.95+0.85+0.30+0.04}_{-0.65-0.68-0.28-0.03}\times 10^{-5}$ in Table~\ref{tab2} is consistent with 
the data $(1.77\pm0.26\pm0.19\pm0.67\pm0.20)\times10^{-5}$ given by LHCb~\cite{prd92-012012}.
While for the other two decay modes, there are apparent inconsistencies for the branching ratios between the PQCD predictions
and the results from Belle and LHCb Collaborations. The Belle's branching fraction~\cite{prd76-012006} for the decay 
$\bar B^0\to D^{*+}_0\pi^-\to D^0\pi^+\pi^-$ is only about $21\%$ of the PQCD prediction in this work, 
other two values from LHCb~\cite{prd92-032002} in Table~\ref{tab1} for this process are some larger, 
but still less than $30\%$ of our result when considering only the central values.
The data for the decay $B^-\to D^{*0}_0K^-\to D^+\pi^-K^-$ selected from LHCb~\cite{prd91-092002} is probably worse
than the $\bar B^0\to D^{*+}_0\pi^-\to D^0\pi^+\pi^-$ case, the branching fraction 
${\mathcal B}=(6.1\pm1.9\pm0.5\pm1.4\pm0.4)\times10^{-6}$ is about one order of magnitude smaller than the predicted value in 
Table~\ref{tab2}.

The $B\to D^*_0(\to D\pi) \pi$ processes can be decomposed in terms of two isospin amplitudes, $A_{1/2}$ and $A_{3/2}$, 
as have been done for the $\bar B^0\to D^{+}\pi^-$ and $B^-\to D^{0}\pi^-$ decays in Ref.~\cite{PRD92-094016}.
With the absolute value $|A_{1/2}/\sqrt{2}A_{3/2}|$ and the relative strong phase
between $A_{1/2}$ and $A_{3/2}$ in~\cite{PRD92-094016}, we have the ratio $R\approx0.59$, 
which is close to the result $R\approx0.54$ from Table~\ref{tab3}, between the branching 
fractions of decays $\bar B^0\to D^{+}\pi^-$ and $B^-\to D^{0}\pi^-$. It is reasonable to expect that the ratio between the branching 
fractions of the decays $\bar B^0\to D^{*+}_0(\to D^0\pi^+)\pi^-$ and $B^-\to D^{*0}_0(\to D^+\pi^-)\pi^-$ is not far from 
the $0.54$. Our results in this work provide $R\approx0.48$ for the corresponding two decays with pion as the 
bachelor particle, while the value for $R$ from the data in Table I is just slightly larger than $0.1$. So one could conclude that the 
data in Table I for the decays $\bar B^0\to D^{*+}_0(\to D^0\pi^+)\pi^-$ and $B^-\to D^{*0}_0(\to D^+\pi^-)\pi^-$ are probably 
inconsistent with the isospin relation.

\begin{figure}[tbp]
\centerline{\epsfxsize=7.0cm \epsffile{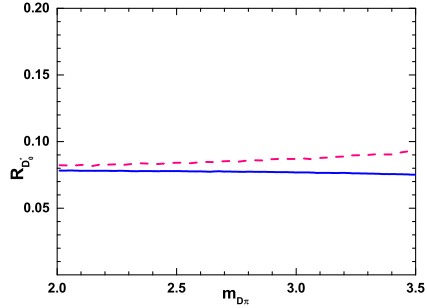}}
\vspace{-0.5cm}
\caption{Energy dependent ratios for the branching fractions between the decays $B^-\to D^{*0}_0K^-\to D^+\pi^-K^-$ and $B^-\to D^{*0}_0\pi^-\to D^+\pi^-\pi^-$ (the solid curve), $\bar B^0\to D^{*+}_0K^-\to D^0\pi^+K^-$ and $\bar B^0\to D^{*+}_0\pi^-\to D^0\pi^+\pi^-$ 
(the dash curve)}
\label{fig-dep}
\vspace{-0.2cm}
\end{figure}

\begin{table}[thb]
\begin{center}
\caption{Data for the concerned decays from {\it Review of Particle Physics}~\cite{PDG-2018} and the ratios
for the related branching fractions }
\label{tab3}  
\begin{tabular}{l l l l r} \hline
 ~~~Mode       &    \;\; \;\;~${\mathcal B}$~      &   ~~~Mode       &    \;\; \;\;~${\mathcal B}$~      &   $R_{D^{(*)}}$~~~~\;  \\  \hline
$B^-\to D^{0}K^-$  & $(3.63\pm0.12)\times10^{-4}$ \;\;~~~~\;\; 
            & $B^-\to D^0\pi^-$  & $(4.68\pm0.13)\times10^{-3}$ \;\; \;\; \;\;     &  $0.078\pm0.003$  \\
 $B^-\to D^*(2007)^0K^-$ & $(3.97^{+0.31}_{-0.28})\times10^{-4}$
            & $B^-\to D^*(2007)^0\pi^-$ & $(4.90\pm0.17)\times10^{-3}$      &  $0.081^{+0.007}_{-0.006}$~~\;  \\  \hline          
$\bar B^0\;\to D^{+}K^-$\;  & $(1.86\pm0.20)\times10^{-4}$ \;\;\;\;        
           &  $\bar B^0\;\to D^{+}\pi^-$    & $(2.52\pm0.13)\times10^{-3}$ \;\;\;\;     &  $0.074\pm0.009$  \\
 $\bar B^0\;\to D^*(2010)^{+}K^-$ & $(2.12\pm0.15)\times10^{-4}$       
           & $\bar B^0\;\to D^*(2010)^{+}\pi^-$ & $(2.74\pm0.13)\times10^{-3}$  &  $0.077\pm0.007$  \\ \hline
\end{tabular}
\end{center}
\vspace{-0.2cm}
\end{table}
For the decay processes $B^-\to D^{*0}_0\pi^-\to D^+\pi^-\pi^-$ and $B^-\to D^{*0}_0K^-\to D^+\pi^-K^-$, we have an 
identical step $D^{*0}_0\to D^+\pi^-$, the difference of the two decay modes originated from the bachelor particles pion and kaon.  
Within the $SU(3)$ flavor symmetry, we have a straightforward ratio $R_{D^{*0}_0}$ for the branching fractions of these two decays as
\begin{eqnarray}
R_{D^{*0}_0}=\frac{{\mathcal{B}(B^-\to D^{*0}_0K^-\to D^+\pi^-K^-)}}{{\mathcal{B}(B^-\to D^{*0}_0\pi^-\to D^+\pi^-\pi^-)}}
\approx \left|\frac{V_{us}}{V_{ud}}\right|^2\cdot \frac{f^2_K}{f^2_\pi}\;,
\label{def-RD0}
\end{eqnarray}
with
\begin{eqnarray}
\left|\frac{V_{us}}{V_{ud}}\right|\frac{f_{K^+}}{f_{\pi^+}}=0.276
\end{eqnarray}
from {\it Review of Particle Physics}~\cite{PDG-2018}, then we have $R_{D^{*0}_0}\approx0.076$. It's easy to obtain a similar ratio 
$R_{D^{*+}_0}\approx R_{D^{*0}_0}$,
\begin{eqnarray}
R_{D^{*+}_0}=\frac{{\mathcal{B}(\bar B^0\to D^{*+}_0K^-\to D^0\pi^+K^-)}}{{\mathcal{B}(\bar B^0\to D^{*+}_0\pi^-\to D^0\pi^+\pi^-)}}
\approx \left|\frac{V_{us}}{V_{ud}}\right|^2\cdot \frac{f^2_K}{f^2_\pi}
\label{def-RDplus}
\end{eqnarray}
for the decay modes $\bar B^0\to D^{*+}_0K^-\to D^0\pi^+K^-$ and $\bar B^0\to D^{*+}_0\pi^-\to D^0\pi^+\pi^-$. 
The energy dependent curves of the $R_{D^{*0}_0}$ and $R_{D^{*+}_0}$ predicted by PQCD are shown in Fig.~\ref{fig-dep}, 
from which one can find that there is little variation for the $R_{D^{*0}_0}$ or $R_{D^{*+}_0}$ as $m_{D\pi}$ runs
from its threshold to $3.5$ GeV.  There are similar patterns for the ratios of the related branching fractions for the decay modes 
including a pseudoscalar $D$ or a vector $D^*$ rather than $D^{*}_0$ as listed in the Table~\ref{tab3}. 
If we accept the average value ${\mathcal B}(B^-\to D^{*0}_0\pi^-)\times{\mathcal B}(D^{*0}_0\to D^+\pi^-)
=(6.4\pm1.4)\times10^{-4}$ in Ref.~\cite{PDG-2018}, the branching fraction ${\mathcal B}=(4.86\pm1.06)\times 10^{-5}$, 
which agree well with the PQCD prediction in Table~\ref{tab2},  
for the decay process $B^-\to D^{*0}_0K^-\to D^+\pi^-K^-$ could be derived from Eq.~(\ref{def-RD0}). 
If we believe the result ${\mathcal B}=(1.77\pm0.77)\times10^{-5}$ given by LHCb~\cite{prd92-012012} for the 
decay $B^0\to D^{*-}_0K^+\to\bar D^0\pi^-K^+$, then the three values listed in Table~\ref{tab1} for the decay
$\bar B^0\to D^{*+}_0\pi^-\to D^0\pi^+\pi^-$ announced by Belle and LHCb are simply not credible when considering 
the Eq~(\ref{def-RDplus}). 
In fact, there is a {\it preliminary result} from the Dalitz plot analysis of the $B^0\to \bar D^0\pi^+\pi^-$ decay processes 
in Ref.~\cite{1007-4464} announced by BaBar as
\begin{eqnarray}
{\mathcal B}(B^0\to D^{*-}_0\pi^+)\times{\mathcal B}(D^{*-}_0\to \bar D^0\pi^-)
 = (2.18\pm0.23\pm0.33\pm1.15\pm0.03)\times 10^{-4}\;.
\end{eqnarray}
This result is consistent with the prediction $2.85^{+1.23+1.05+0.33+0.06}_{-0.80-0.81-0.31-0.05}\times 10^{-4}$ within errors.

To sum up, we studied the quasi-two-body decays $B^-\to D^{*0}_0\pi^-\to D^+\pi^-\pi^-$, 
$\bar B^0\to D^{*+}_0\pi^-\to D^0\pi^+\pi^-$, $B^-\to D^{*0}_0K^-\to D^+\pi^-K^-$ and 
$\bar B^0\to D^{*+}_0K^-\to D^0\pi^+K^-$ in the PQCD approach in this work. 
The branching fraction predicted by PQCD for the decay process $B^-\to D^{*0}_0\pi^-\to D^+\pi^-\pi^-$ agree well 
with the data from Belle, BaBar and LHCb Collaborations. The result for the $\bar B^0\to D^{*+}_0K^-\to D^0\pi^+K^-$ in this 
work is consistent with the data $(1.77\pm0.26\pm0.19\pm0.67\pm0.20)\times10^{-5}$ given by LHCb.
For the other two decays, we analyzed the experimental results using the ratio relations 
between the decay branching fractions including $D^{*0}_0$ or $D^{*+}_0$. 
From $R_{D^{*0}_0}$ and $R_{D^{*+}_0}$, we argued that the experimental results for the decays 
$\bar B^0\to D^{*+}_0\pi^-\to D^0\pi^+\pi^-$ and $B^-\to D^{*0}_0K^-\to D^+\pi^-K^-$ are probably questionable.  
The PQCD predictions for these two decay modes, in this work, are
$2.85^{+1.23}_{-0.80} (\omega_B) ^{+1.05}_{-0.81}(\omega_{D\pi})^{+0.33}_{-0.31}(a_{D\pi})
                  ^{+0.06}_{-0.05}(\Gamma_{D^{*+}_0}) \times 10^{-4}$ and 
$4.65^{+1.89}_{-1.30} (\omega_B) ^{+1.51}_{-1.24}(\omega_{D\pi})^{+0.40}_{-0.38}(a_{D\pi})
                  ^{+0.22}_{-0.18}(\Gamma_{D^{*0}_0}) \times 10^{-5}$, respectively.
We concluded that the available experimental results for the four decays including $D^\ast_0(2400)$ are not in agreement with 
the isospin relation and $SU(3)$ flavor symmetry. 

\begin{acknowledgments}
This work was supported in part by National Science Foundation of China under Grant No. 11547038.
\end{acknowledgments}

\appendix

\section{Decay amplitudes}  

The concerned quasi-two-body decay amplitudes are given, in the PQCD approach, by
\begin{eqnarray}
{\mathcal A}\big(B^-\to \pi^-[ D_0^{*0}\to]D^+\pi^-\big)&=&\frac{G_F}{\sqrt2}V_{cb}V^*_{ud}
\big[\big(\frac{c_1}{3}+c_2\big)F_{TD_0^*} +c_1M_{TD_0^*} +\big(c_1+\frac{c_2}{3}\big) F_{T\pi} + c_2 M_{T\pi}\big]\;, \\
{\mathcal A}\big(B^-\to K^-[D_0^{*0}\to]D^+\pi^-\big)&=&\frac{G_F}{\sqrt2}V_{cb}V^*_{us}
\big[\big(\frac{c_1}{3}+c_2\big)F_{TD_0^*} +c_1M_{TD_0^*} +\big(c_1+\frac{c_2}{3}\big) F_{TK} + c_2 M_{TK}\big]\;, \\
{\mathcal A}\big(\bar B^0\to \pi^-[ D_0^{*+}\to] D^0\pi^+\big)&=&\frac{G_F}{\sqrt2}V_{cb}V^*_{ud}
\big[\big(\frac{c_1}{3}+c_2\big)F_{TD_0^*} +c_1M_{TD_0^*} +\big(c_1+\frac{c_2}{3}\big) F_{A\pi} + c_2 M_{A\pi}\big]\;, \\
{\mathcal A}\big(\bar B^0\to K^-[ D_0^{*+}\to] D^0\pi^+\big)&=&\frac{G_F}{\sqrt2}V_{cb}V^*_{us}
\big[\big(\frac{c_1}{3}+c_2\big)F_{TD_0^*} +c_1M_{TD_0^*}\big]\;, 
\end{eqnarray}
in which $G_F$ is the Fermi coupling constant, $V$'s are the CKM matrix elements.
And it should be understood that the Wilson coefficients $c_1$ and $c_2$ appear in convolutions in momentum fractions and 
impact parameters $b$.

The amplitudes from Fig.~1 are written as
\begin{eqnarray}
F_{TD_0^*} &=& 8\pi C_F m^4_B f_{\pi(K)} (\eta-1)\int dx_B dz\int b_B db_B b db \phi_B(x_B,b_B)\phi_{D\pi}(z,b,s) \nonumber\\
&\times&\big\{\big[\sqrt{\eta}(2z-1)-1-z \big] E^{(1)}_{a}(t^{(1)}_{e})h(x_B,z,b_B,b)
+\left(2\sqrt{\eta}(r_c-1)+\eta-r_c \right) E^{(2)}_{a}(t^{(2)}_{e})h(z,x_B,b,b_B) \big\}, \\
M_{TD_0^*} &=& 32\pi C_F m^4_B/\sqrt{2N_c}(\eta-1) \int dx_B dz dx_3\int b_B db_B b_3 db_3\phi_B(x_B,b_B)
\phi_{D\pi}(z,b,s)\phi^A \nonumber\\
&\times& \big\{ \left[\eta\left(1-z-x_3\right)+z\sqrt{\eta}+(x_B+x_3-1)\right]E_{b}(t^{(1)}_{b})h^{(1)}_{b}(x_i,b_i)\nonumber\\
&+&\left[x_3\left(1-\eta\right)+z\left(1-\sqrt{\eta}\right) -x_B \right]E_{b}(t^{(2)}_{b})h^{(2)}_{b}(x_i,b_i) \big\}\;, \\
F_{T\pi(K)}&=& 8\pi C_F m^4_B F_{D\pi}(s)  \int dx_B dx_3\int b_B db_B b_3 db_3 \phi_B(x_B,b_B)\nonumber\\
&\times &\big\{\big[\phi^A(1-\eta)(x_3(\eta-1)-1)
 -r_0[\phi^P(\eta+1+2(\eta-1)x_3)+\phi^T(\eta-1)(2x_3-1)]\big]E^{(1)}_{c}(t^{(1)}_{i}) \nonumber\\
&\times&  h(x_B,x_3(1-\eta),b_B,b_3) +\left[2r_0\phi^P(\eta(1+ x_B)-1)+\eta(\eta-1) x_B\phi^A\right]E^{(2)}_{c}(t^{(2)}_{i})
h(x_3,x_B(1-\eta),b_3,b_B)\big\}\;,\\
M_{T\pi(K)} &=& 32\pi C_F m^4_B/\sqrt{2N_c} \int dx_B dz dx_3\int b_B db_B b db\phi_B(x_B,b_B)\phi_{D\pi}(z,b,s) \nonumber\\
&\times&\big\{\big[  \phi^A(1-\eta)(\sqrt\eta r_c+ (1+\eta)(1-x_B-z)) +r_0\phi^P(\eta(x_B+z+x_3-2)-4\sqrt\eta r_c-x_3)  \nonumber\\
&+& r_0\phi^T(\eta (x_B+z-x_3)+x_3)\big] E_{d}(t^{(1)}_{d})h^{(1)}_{d}(x_i,b_i)+\big[(\eta-1)[z-x_B+(1-\eta)x_3]\phi^A\nonumber\\
&+&r_0x_3(1-\eta)(\phi^P+\phi^T)+r_0\eta(x_B-z)(\phi^T-\phi^P)\big]E_{d}(t^{(2)}_{d})h^{(2)}_{d}(x_i,b_i)\big\}\;,\\
F_{A\pi} &=& 8\pi C_F m^4_B f_B \int dz dx_3\int b db b_3 db_3 \phi_{D\pi}(z,b,s))\nonumber\\
&\times&\big\{ \big[(\eta-1)[(1+2\sqrt{\eta}r_c+(\eta-1)x_3)\phi^A +r_0(r_c+2x_3\sqrt{\eta})\phi^T]  
-r_0[(1+\eta)r_c+2\sqrt{\eta}(x_3(\eta-1)+2)]\phi^P\big] \nonumber\\
&\times& E^{(1)}_{e}(t^{(1)}_{a})h_{a}(z,x_3(1-\eta),b,b_3)+ \left[(1-\eta)z\phi^A+2r_0 \sqrt{\eta}(1-\eta+z)\phi^P\right]  
E^{(2)}_{e}(t^{(2)}_{a})h_{a}(x_3,z(1-\eta),b_3,b) \big\},\\
M_{A\pi} &=& 32\pi C_F m^4_B/\sqrt{2N_c} \int dx_B dz dx_3\int b_B db_B b_3 db_3\phi_B(x_B,b_B)\phi_{D\pi}(z,b,s)  \nonumber\\
&\times& \big\{ \big[ (\eta-1) ( \eta (x_B+z-1)+z+x_B)  \phi^A 
+r_0 \sqrt \eta [(\eta-1)(1-x_3)(\phi^P+\phi^T) + (z+x_B)(\phi^T-\phi^P)-2\phi^P] \big]\quad\nonumber\\
&\times& E_{f}(t^{(1)}_{f})h^{(1)}_{f}(x_i,b_i) +\big[ (1-\eta) [\eta (z-x_B)+(1-x_3)(1-\eta)] \phi^A +r_0 \sqrt \eta [ (\eta-1)(x_3-1) \nonumber\\
&\times&(\phi^P-\phi^T)   + (z-x_B)(\phi^P+\phi^T)    ]   \big]E_{f}(t^{(2)}_{f})h^{(2)}_{f}(x_i,b_i) \big\},
\end{eqnarray}

The evolution factors in the above factorization formulas are given by
\begin{eqnarray}
E^{(1)}_a(t)&=&\alpha_s(t){\rm exp}[-S_B(t)-S_C(t)] S_t(z)\;, \quad ~E^{(2)}_a(t)=\alpha_s(t){\rm exp}[-S_B(t)-S_C(t)]  S_t(x_B)\;, \\
E_b(t) &=& \alpha_s(t){\rm exp}[-S_B(t)-S_C(t)-S_P(t)]|_{b=b_B}\;,\\
E^{(1)}_c(t)&=&\alpha_s(t){\rm exp}[-S_B(t)-S_P(t)] S_t(x_3)\;, \quad E^{(2)}_c(t)=\alpha_s(t){\rm exp}[-S_B(t)-S_P(t)]  S_t(x_B)\;, \\
E_d(t) &=& \alpha_s(t){\rm exp}[-S_B(t)-S_C(t)-S_P(t)]|_{b_3=b_B}\;,\\
E^{(1)}_e(t)&=&\alpha_s(t){\rm exp}[-S_C(t)-S_P(t)] S_t(x_3)\;, \quad E^{(2)}_e(t)=\alpha_s(t){\rm exp}[-S_C(t)-S_P(t)]  S_t(z)\;, \\
E_f(t) &=& \alpha_s(t){\rm exp}[-S_B(t)-S_C(t)-S_P(t)]|_{b_3=b}\;.
\end{eqnarray}
in which $S_{(B,C,P)}(t)$ are in the Appendix of~\cite{prd78-014018}, the hard functions $h, h_a, h^{(1,2)}_{(b,d,f)}$ and the 
hard scales $t^{(1,2)}_{(e,b,i,d,a,f)}$ have their explicit expressions in the Ref.~\cite{prd78-014018}. We need to stress that, 
because of the different definitions of the momenta for the initial and final states, the concerned expressions in~\cite{prd78-014018}
could be employed in this work only after the replacements $\{x_1\to x_B, b_1\to b_B, x_2\to z, b_2\to b, r^2\to\eta\}$.
The parameter $c$ in the Eq.~(A1) of~\cite{prd78-014018} is adopt to be 0.4 in this work according to the 
Refs.~\cite{prd65-014007,prd80-074024}.


\end{document}